# Generalized Polya Urn for Time-varying Dirichlet Process Mixtures

**François Caron**
Dept. of Computer Science
University of British Columbia
Vancouver, Canada
caronfr@cs.ubc.ca

**Manuel Davy**
CNRS/INRIA Futurs Sequel
Ecole Centrale de Lille
Villeneuve d'Ascq, France
manuel.davy@inria.fr

**Arnaud Doucet**
Depts. of Computer Science & Statistics
University of British Columbia
Vancouver, Canada
arnaud@cs.ubc.ca

## Abstract

Dirichlet Process Mixtures (DPMs) are a popular class of statistical models to perform density estimation and clustering. However, when the data available have a distribution evolving over time, such models are inadequate. We introduce here a class of time-varying DPMs which ensures that at each time step the random distribution follows a DPM model. Our model relies on an intuitive and simple generalized Polya urn scheme. Inference is performed using Markov chain Monte Carlo and Sequential Monte Carlo. We demonstrate our model on various applications.

## 1 INTRODUCTION

### 1.1 Background

Dirichlet Process Mixtures (DPMs) are a generalization of finite mixture models to infinite mixtures. They have become very popular over the past few years in machine learning and statistics to perform clustering and density estimation. However, there are many situations where we cannot assume that the distribution of the observations is fixed and instead this latter evolves over time. For example, in a clustering application, the number of clusters and the locations of these clusters may change over time. This article introduces a novel class of time-varying first-order stationary DPM; that is processes which have marginals following the same DPM. We first briefly recall standard results about DPMs.

Let $t = 1, 2, ...$ denote a discrete-time index. For any generic sequence $\{x_m\}$, we define $x_{k:l} = (x_k, x_{k+1}, \ldots, x_l)$. For ease of presentation, we assume that we receive a fixed number $n$ of observations at each time $t$ denoted $\mathbf{z}_t = \mathbf{z}_{1:n,t}$ which are independent and identically distributed (iid) samples from

$$F_t(\cdot) = \int_{\mathcal{Y}} f(\cdot|\mathbf{y}) d\mathbb{G}_t(\mathbf{y}) \tag{1}$$

where $f(\cdot|\mathbf{y})$ is the mixed pdf and $\mathbb{G}_t$ is the mixing distribution distributed according to a Dirichlet Process (DP)

$$\mathbb{G}_t \sim DP(\theta, \mathbb{G}_0) \tag{2}$$

where $\mathbb{G}_0$ is a base probability measure and $\theta > 0$ is the scale parameter. The random measure $\mathbb{G}_t$ satisfies the following *stick-breaking* representation [18]

$$\mathbb{G}_t = \sum_{k=1}^{\infty} V_{k,t} \delta_{U_{k,t}} \tag{3}$$

where $V_{k,t} = \beta_{k,t} \prod_{j=1}^{k-1}(1 - \beta_{j,t})$, $\beta_{k,t} \overset{\text{iid}}{\sim} \mathcal{B}(1, \theta)$ and $U_{k,t} \overset{\text{iid}}{\sim} \mathbb{G}_0$. From (1), we have equivalently the following hierarchical model

$$\mathbf{y}_{k,t}|\mathbb{G}_t \overset{\text{iid}}{\sim} \mathbb{G}_t, \tag{4}$$

$$\mathbf{z}_{k,t}|\mathbf{y}_{k,t} \overset{\text{iid}}{\sim} f(\cdot|\mathbf{y}_{k,t}). \tag{5}$$

We can also reformulate the DPM by integrating out the mixing measure $\mathbb{G}_t$ and introducing allocation variables $\mathbf{c}_t = c_{1:n,t}$ such that for any $j \in \mathcal{J}(\mathbf{c}_t)$, where $\mathcal{J}(\mathbf{c}_t)$ is the set of unique values in $\mathbf{c}_t$, we have

$$U_{j,t} \overset{\text{iid}}{\sim} \mathbb{G}_0, \tag{6}$$

and

$$\mathbf{z}_{k,t}|U_{c_{k,t}} \sim f(\cdot|U_{c_{k,t},t}). \tag{7}$$

For convenience, we label here the clusters by their order of appearance. We set $c_{1,1} = 1$, $K_1 = 1$ and $\mathbf{m}_1^1 = m_{1:K_1,1}^1$ a vector of size $K_1$. Then, at time $t = 1$, for $k = 2, \ldots, n$ we have the following Polya urn model [2]

w.p. $\frac{m_{1,1}^1}{k-1+\theta}$, $i \in \{1, \ldots, K_1\}$ set $c_{k,1}^1 = i$, $m_{i,1}^1 = m_{i,1}^1 + 1$,

w.p. $\frac{\theta}{k-1+\theta}$, set $K_1 = K_1 + 1$, $c_{k,1}^1 = K_1$, $m_{K_1,1}^1 = 1$,

where the notation 'w.p.' stands for 'with probability'.

The sequence $c_{1,t}, \ldots, c_{n,t}$ is exchangeable and induces a random *partition of* $n$, that is an unordered collection of $k \leq n$ positive integers with sum $n$ or, equivalently, a random allocation of $n$ **unlabeled** balls into some random number of **unlabeled** boxes (materialized by a color for example); each box containing at least one ball. One common way to code a partition of $n$ is by the number of terms of various sizes; that is the *vector of counts* $(a_1, \ldots, a_n)$ where $\sum_{j=1}^n a_j = k$ and $\sum_{j=1}^n j a_j = n$. $a_1$ is the number of terms of size 1, $a_2$ is the number of terms of size 2, etc. Following [1], we say that $\mathbf{c}_t \in C(a_{1:n})$ if there are $a_1$ distinct values of $\mathbf{c}_t$ that occur only once, $a_2$ that occur twice, etc. One can show that $\Pr(\mathbf{c}_t \in C(a_{1:n})) = P_n(a_{1:n})$ is given by the Ewens Sampling Formula (ESF)

$$P_n(a_{1:n}) = \frac{n!}{\prod_{i=1}^n (\theta + i - 1)} \prod_{i=1}^n \frac{\theta^{a_i}}{i^{a_i} a_i!} \qquad (8)$$

In this paper, we introduce a statistical model to introduce dependencies between the distributions $\{F_t\}$ and mixing distributions $\{\mathbb{G}_t\}$ while preserving (1) and (2) at any time $t$. Various constructions have already been presented in the literature that we briefly review below.

### 1.2 Literature review

Several authors have considered previously the problems of defining dependent DP for time series or spatial modelling.

In an early contribution, [5] introduced dependencies between distributions by defining a parametric model on the base distribution $\mathbb{G}_{0,s}$ dependent on the covariate $s$ and $\mathbb{G}_s \sim DP(\theta, \mathbb{G}_{0,s})$. This approach is different from ours as we introduce dependencies directly on two successive mixing distributions while $\mathbb{G}_0$ is fixed.

The great majority of recent papers use the stick-breaking representation (3) to introduce dependencies. Under this representation, a realization of a DP is represented by two (infinite dimensional) vectors of weights $V_{1:\infty,s}$ and clusters locations $U_{1:\infty,s}$. Dependency with respect to a covariate $s$ is introduced on $V_{1:\infty,s}$ in [9] and on $U_{1:\infty,s}$ in [13], [10], [8]. An alternative approach is to consider the mixing distribution to be a convex combination of independent random probability measures sampled from a DP. The dependency is then introduced through the weighting coefficients; e.g. [14], [7].

Although these previous approaches have merits, we believe that it is possible to build more intuitive models based on Polya urn-type schemes. We are aware of a technical report [22] proposing a time-varying Polya urn model but this model does not marginally preserve a DP. The only model we know of which satisfies this property is presented in [21] based on the methodology of [16]. The authors define a joint distribution $p(\mathbb{G}_1, \mathbb{G}_2)$ such that $\mathbb{G}_1$ and $\mathbb{G}_2$ are marginally $DP(\theta, \mathbb{G}_0)$ by introducing $m$ artificial auxiliary variables $\mathbf{w}_i \overset{\text{iid}}{\sim} \mathbb{G}_1$ and then $\mathbb{G}_2 | \mathbf{w}_{1:m} \sim DP(\theta + n, \frac{\theta \mathbb{G}_0 + \sum_{i=1}^m \delta_{\mathbf{w}_i}}{\theta + m})$. An extension to time series is discussed in [19]. One important drawback of this approach is that it requires introducing a very large number $m$ of auxiliary variables to model strongly dependent distributions. When inference is performed, these auxiliary variables need to be inferred from the data and the resulting algorithm is very computationally intensive.

### 1.3 Contributions

The model developed in this paper is based on a Polya urn representation of the DP and does not rely on any artificial auxiliary variable. To obtain a first-order stationary DP using such an approach, we need to ensure that any time $t$

(**A**). The sequence $\mathbf{c}_t$ induces a random partition distributed according to the ESF.

(**B**). For $j \in \mathbf{c}_t$, the $U_{j,t}$s are i.i.d. from $\mathbb{G}_0$.

The main contribution of this paper consists of defining models satisfying (**A**) using a generalized Polya urn prediction rule and the consistence properties under specific deletion procedures of the ESF, see Kingman [12]. Ensuring (**B**) can be performed using quite standard methods from the time series literature; e.g. [11]. Moreover our model enjoys the following desirable property:

(**C**). There exists an hyperparameter $0 \leq \rho \leq 1$ to tune the 'closeness' between the random partitions induced by $\mathbf{c}_t$ and $\mathbf{c}_{t+1}$ with these partitions being statistically independent if $\rho = 0$ and being 'close' as $\rho$ increases to 1. In the limiting case where $\rho = 1$ we have a static DP model.

Our model allows us to move both the cluster locations and their weights. Furthermore, it relies on a simple and intuitive birth/death procedure described further. By using a Polya urn approach, the model is defined on the space of equivalence labels, i.e. the labelling of the class to which each data belongs is irrelevant. From a computational point of view, it is usually easier to design efficient Markov chain Monte Carlo (MCMC) when making inference based on this representation compared to the stick-breaking representation, see [17] for a discussion.

## 2 STATIONARY DPM MODELS

We first address here the points (**A**) and (**C**) by describing two first order stationary partition models which can be easily combined if necessary then we discuss (**B**).

### 2.1 Stationary random partition Models

The main idea behind these models consists at each time step $t$ of

• deleting randomly a subset of the allocations variables sampled from time 1 to $t-1$ which had survived the previous $t-1$ deletion steps,

• sampling $n$ new allocation variables corresponding to the $n$ observations $\mathbf{z}_t$.

For any $t \geq 2$, we have generated the allocation variables $\mathbf{c}_{1:t-1}$ corresponding to $\mathbf{z}_{1:t-1}$ from time 1 to $t-1$. We denote by $\mathbf{c}_{1:t-1}^{t-1}$ (resp. $\mathbf{c}_{1:t-1}^{t}$) the subset of $\mathbf{c}_{1:t-1}$ corresponding to variables having survived the deletion steps from time 1 to $t-1$ (resp. from time 1 to $t$). Let $K_{t-1}$ be the number of clusters created from time 1 to $t-1$, we denote by $\mathbf{m}_{t-1}^{t-1}$ (resp. $\mathbf{m}_{t-1}^{t}$) the vector of size $K_{t-1}$ containing the size of the boxes associated to $\mathbf{c}_{1:t-1}^{t-1}$ (resp. $\mathbf{c}_{1:t-1}^{t}$). Hence, these vectors have zero entries corresponding to 'dead' clusters. The introduction of $\mathbf{m}_{t-1}^{t-1}$ and $\mathbf{m}_{t-1}^{t}$ simplify the presentation of the procedure but note that, from a practical point of view, there is obviously no need to store these vectors of increasing dimension. It is only necessary to store the size of the non-empty boxes and their associated labels.

At time 1, we just generate $\mathbf{c}_1$ according to a standard Polya urn described in the introduction. At time $t \geq 2$ we have $\mathbf{c}_{1:t-1}^{t-1} = \left(\mathbf{c}_{1:t-2}^{t-1}, \mathbf{c}_{t-1}\right)$ and we sample $\mathbf{c}_{1:t}^{t} = \left(\mathbf{c}_{1:t-1}^{t}, \mathbf{c}_t\right)$ as follows. We first obtain $\mathbf{c}_{1:t-1}^{t}$ by deleting a random number of balls from $\mathbf{c}_{1:t-1}^{t-1}$ according to one of the following rules.

• **Uniform deletion**: delete each allocation variable in $\mathbf{c}_{1:t-1}^{t-1}$ with probability $1-\rho$ where $0 \leq \rho \leq 1$. This is statistically equivalent to sample a number $r$ from a binomial distribution $\mathcal{B}\text{in}(\sum_k m_{k,t-1}^{t-1}, 1-\rho)$ and then to remove $r$ items uniformly from $\mathbf{c}_{1:t-1}^{t-1}$ to obtain $\mathbf{c}_{1:t-1}^{t}$.

• **Size-biased deletion**: we compute the following discrete probability distribution over the set of non-empty boxes $\frac{m_{k,t-1}^{t-1}}{\sum_i m_{i,t-1}^{t-1}}$, sample a realization from this distribution and delete the corresponding box to obtain $\mathbf{c}_{1:t-1}^{t}$.

It is also possible to consider any mixture and composition of these strategies. For example, we can pick w.p. $\alpha$ the uniform deletion strategy and w.p. $1-\alpha$ the size-biased deletion strategy or perform one uniform deletion followed by one size-biased deletion or two size-biased deletions etc. The size-biased deletion allows us to model large potential jumps in the distributions of the observations. Finally, after this deletion step, we sample the allocation variables $\mathbf{c}_t$ according to a standard Polya urn scheme based on the surviving allocation variables $\mathbf{c}_{1:t-1}^{t}$. To summarize, the generalized Polya urn scheme proceeds as follows where $\mathcal{I}(\mathbf{m}_t^t)$ denotes the indices corresponding to the non-zero entries of $\mathbf{m}_t^t$.

---

**Generalized Polya Urn**

---

<u>*At time* $t=1$</u>

• Set $c_{1,1}^1 = 1$, $m_{1,1}^1 = 1$ and $K_1 = 1$.

• For $k = 2, ..., n$

w.p. $\frac{m_{i,1}^1}{k-1+\theta}$, $i \in \{1, \ldots, K_1\}$, set $c_{k,1}^1 = i$, $m_{i,1}^1 = m_{i,1}^1 + 1$,

w.p. $\frac{\theta}{k-1+\theta}$, set $K_1 = K_1 + 1$, $c_{k,1}^1 = K_1$, $m_{K_1,1}^1 = 1$.

<u>*At time* $t \geq 2$</u>

• Kill randomly a subset of $\mathbf{c}_{1:t-1}^{t-1}$ using uniform and/or size-biased deletion to obtain $\mathbf{c}_{1:t-1}^{t}$ (hence $\mathbf{m}_{t-1}^{t}$) and set $\mathbf{m}_t^t = \mathbf{m}_{t-1}^t$, $K_t = K_{t-1}$.

• For $k = 1, ..., n$

w.p. $\frac{m_{i,t}^t}{\sum_i m_{i,t}^t + \theta}$, $i \in \mathcal{I}(\mathbf{m}_t^t)$, set $c_{k,t}^t = i$, $m_{i,t}^t = m_{i,t}^t + 1$,

w.p. $\frac{\theta}{\sum_i m_{i,t}^t + \theta}$, set $K_t = K_t + 1$, $c_{k,t}^t = K_t$, $m_{K_t,t}^t = 1$.

---

Our main result is that $\mathbf{c}_t$ satisfies (**A**). It is a consequence of the remarkable consistence properties under deletion of the ESF which have been first established by Kingman [12] and are also mentioned for example in [20].

**Proposition**. At any time $t \geq 1$, $\mathbf{c}_t$ induces a random partition distributed according to the ESF.

*Proof.* We prove by induction a stronger result; that is $\mathbf{c}_{1:t}^t$ induces a random partition following the ESF. At time 1, this is trivially true as $\mathbf{c}_1^1 = \mathbf{c}_1$ is generated according to a standard Polya urn. Assume it is true at time $t-1$, then the specific deletion steps we have proposed ensure that $\mathbf{c}_{1:t-1}^t$ also induces a random partition following the ESF thanks to the results in [12, pp. 3 and 5]. Finally, as $\mathbf{c}_t$ is sampled according to a standard Polya urn scheme based on the surviving allocation variables $\mathbf{c}_{1:t-1}^t$ then $\mathbf{c}_{1:t}^t$ indeed induces by construction a random partition following the ESF.

Now thanks to exchangeability, it implies that $\mathbf{c}_t$ also induces a random partition distributed according to the ESF.■

Clearly if we only use uniform deletion (i.e. $\alpha = 1$), then we also see that (**C**) is satisfied.

**Remark**. Instead of the uniform deletion models proposed above, a $r$-order Markov (sliding window) model could also be defined where the allocation variables $\mathbf{c}_{t-r-1}$ are deterministically killed at each time $t$. In this case, the vector $\mathbf{m}_{t-1}^{t}$ contains the sizes of the boxes associated to $\mathbf{c}_{t-r:t-1}$. This model also induces a random partition following the ESF.

### 2.2 Correlation Structure

Clearly the sequence $\{\mathbf{c}_t\}$ is not Markovian but $\{\mathbf{c}_{1:t}^{t}\}$ and $\left\{\mathbf{c}_{1:t-1}^{t}\right\}$ and the associated vectors $\{\mathbf{m}_{1:t}^{t}\}$ and $\left\{\mathbf{m}_{1:t-1}^{t}\right\}$ are Markovian. We can compute analytically the transition probabilities for these processes but the resulting expressions are quite complex. However it can be shown easily for example that for the uniform deletion model we have for $k \in \mathcal{I}(\mathbf{m}_{t-1}^{t})$

$$\begin{aligned}\mathbb{E}\left[m_{k,t}^{t+1}|\mathbf{m}_{t-1}^{t}\right] &= \mathbb{E}\left[\mathbb{E}\left[m_{k,t}^{t+1}\middle| m_{k,t}^{t}\right]|\mathbf{m}_{t-1}^{t}\right] \\ &= \rho\left(m_{k,t-1}^{t} + n\frac{m_{k,t-1}^{t}}{\theta + \sum_k m_{k,t-1}^{t}}\right)\end{aligned}$$

and

$$\mathbb{E}[\sum_{k\notin\mathcal{I}(\mathbf{m}_{t-1}^{t})} m_{k,t}^{t+1}|\mathbf{m}_{t-1}^{t}] = \frac{\rho n\theta}{\theta + \sum_k m_{k,t-1}^{t}}.$$

It can also be shown that $\mathbb{G}_t$ is asymptotically a second order stationary process, that is $cov(\int \varphi(\mathbf{y})\,\mathbb{G}_t(d\mathbf{y}), \int \varphi(\mathbf{y})\,\mathbb{G}_{t+\tau}(d\mathbf{y}))$ is a function of $|\tau|$ for large $t$. We display a Monte Carlo estimate of $corr\left(\int \mathbf{y}\mathbb{G}_t(d\mathbf{y}), \int \mathbf{y}\mathbb{G}_{t+\tau}(d\mathbf{y})\right)$ when $\mathbb{G}_0$ is a standard normal distribution for different values of $\rho$ and $\theta$ in Fig. 1. The correlations decrease faster as $\rho$ decreases as expected.

### 2.3 Stationary Models for Clusters Locations and Time-Varying DPM

In the previous subsections, we have presented some stationary models for allocation variables. To obtain a first order stationary DPM process, we also need to ensure point (**B**). This can be easily achieved if for $k \in \mathcal{I}(\mathbf{m}_{t}^{t})$

$$U_{k,t} \sim \begin{cases} p\left(U_{k,t}|U_{k,t-1}\right) & \text{if } k \in \mathcal{I}(\mathbf{m}_{t-1}^{t}) \\ \mathbb{G}_0 & \text{otherwise} \end{cases}$$

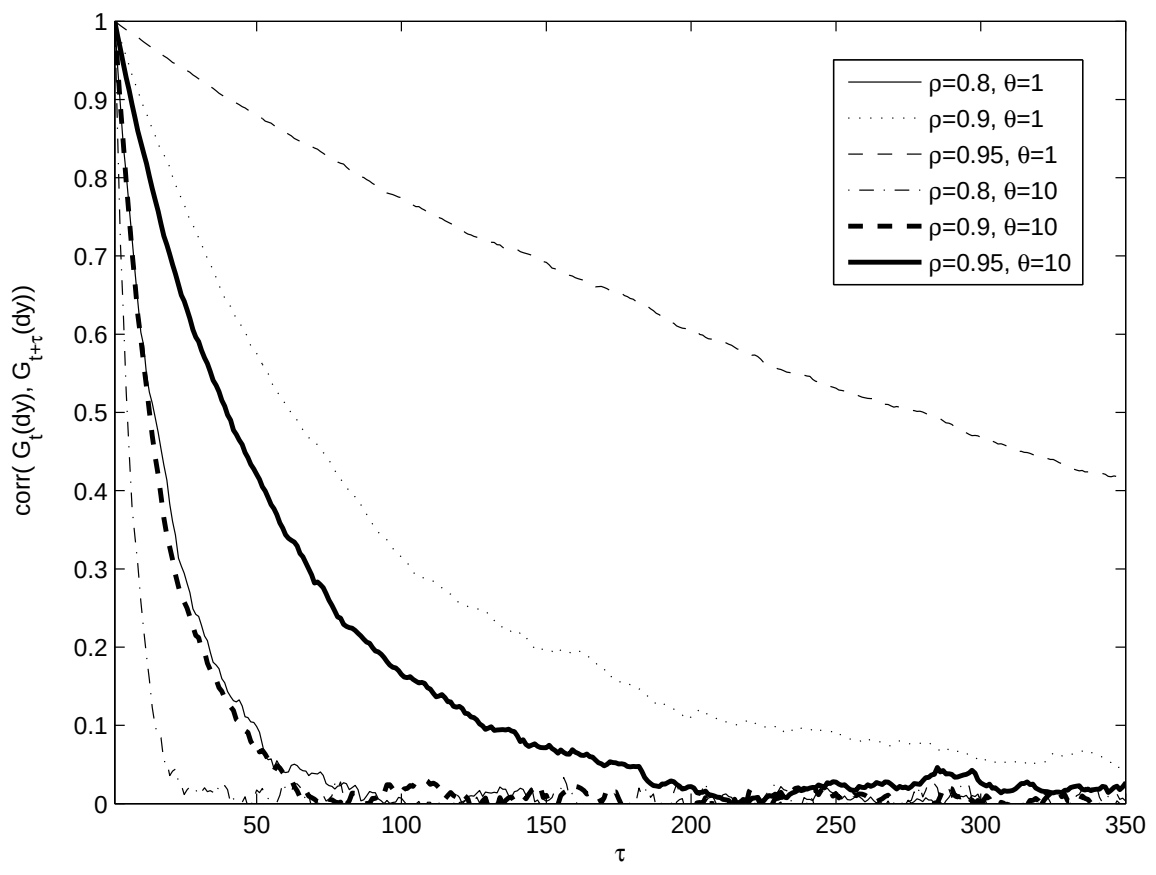

Figure 1: $corr\left(\int \mathbf{y}\mathbb{G}_t(d\mathbf{y}), \int \mathbf{y}\mathbb{G}_{t+\tau}(d\mathbf{y})\right)$ approximated by Monte Carlo simulations in function of $\tau$ for different values of $\rho$ and $\theta$.

where $\mathbb{G}_0$ is the invariant distribution of the transition kernel $p\left(\cdot|\cdot\right)$, i.e.

$$\int \mathbb{G}_0\left(U_{k,t-1}\right) p\left(U_{k,t}|U_{k,t-1}\right) dU_{k,t-1} = \mathbb{G}_0\left(U_{k,t}\right).$$

In the time series literature, many approaches are available to build such transition kernels based on copulas [11] or auxiliary variables [16]. Note that applied to 'standard' time series and not to DP, the approach in [16] does not typically suffer from the problem outlined in Section 1.2.

Combining the stationary DP and cluster location models, we can summarized the model by the following graphical model in Fig. 2. It can also be summarized by the metaphor of the Chinese restaurant, see Fig. 3.

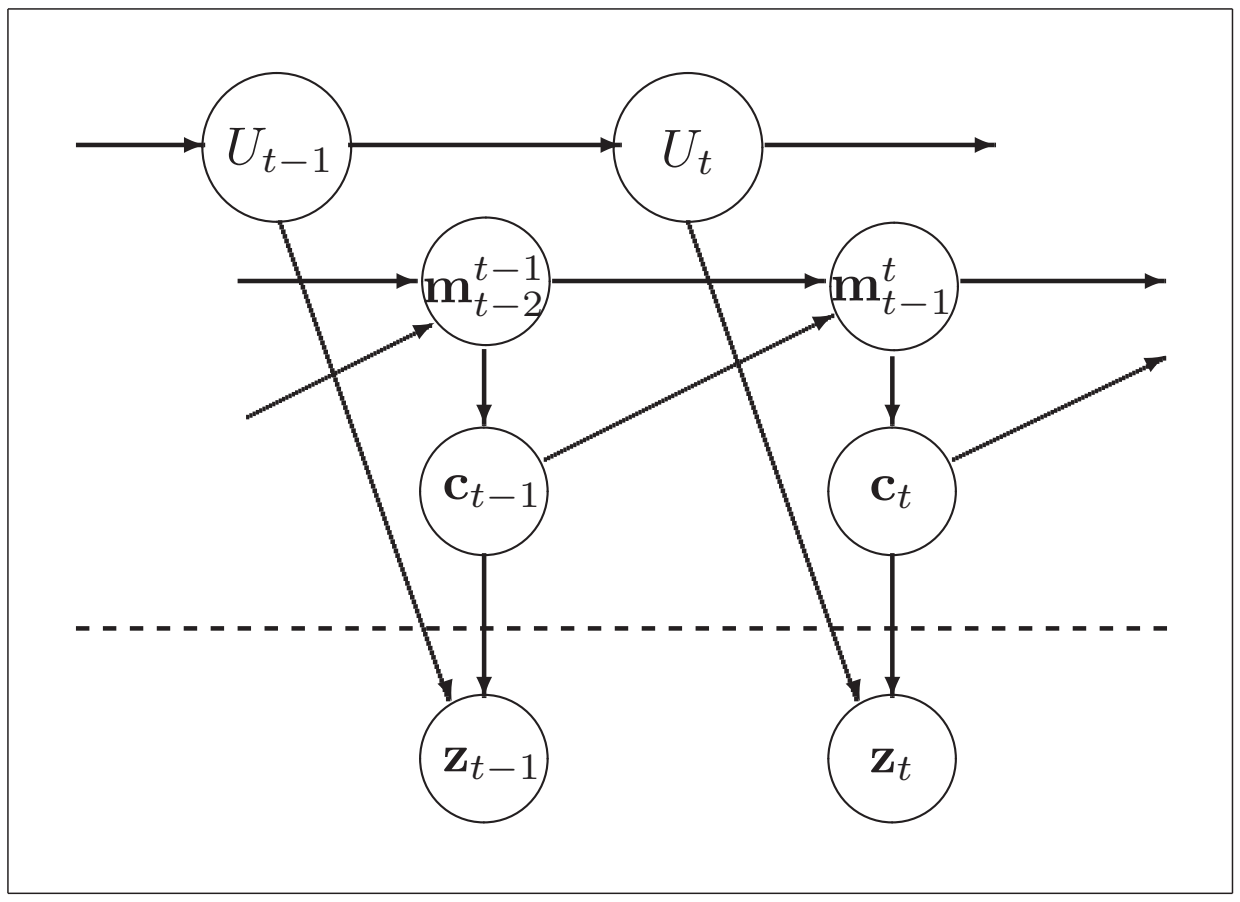

Figure 2: Graphical model of the TVDPM.

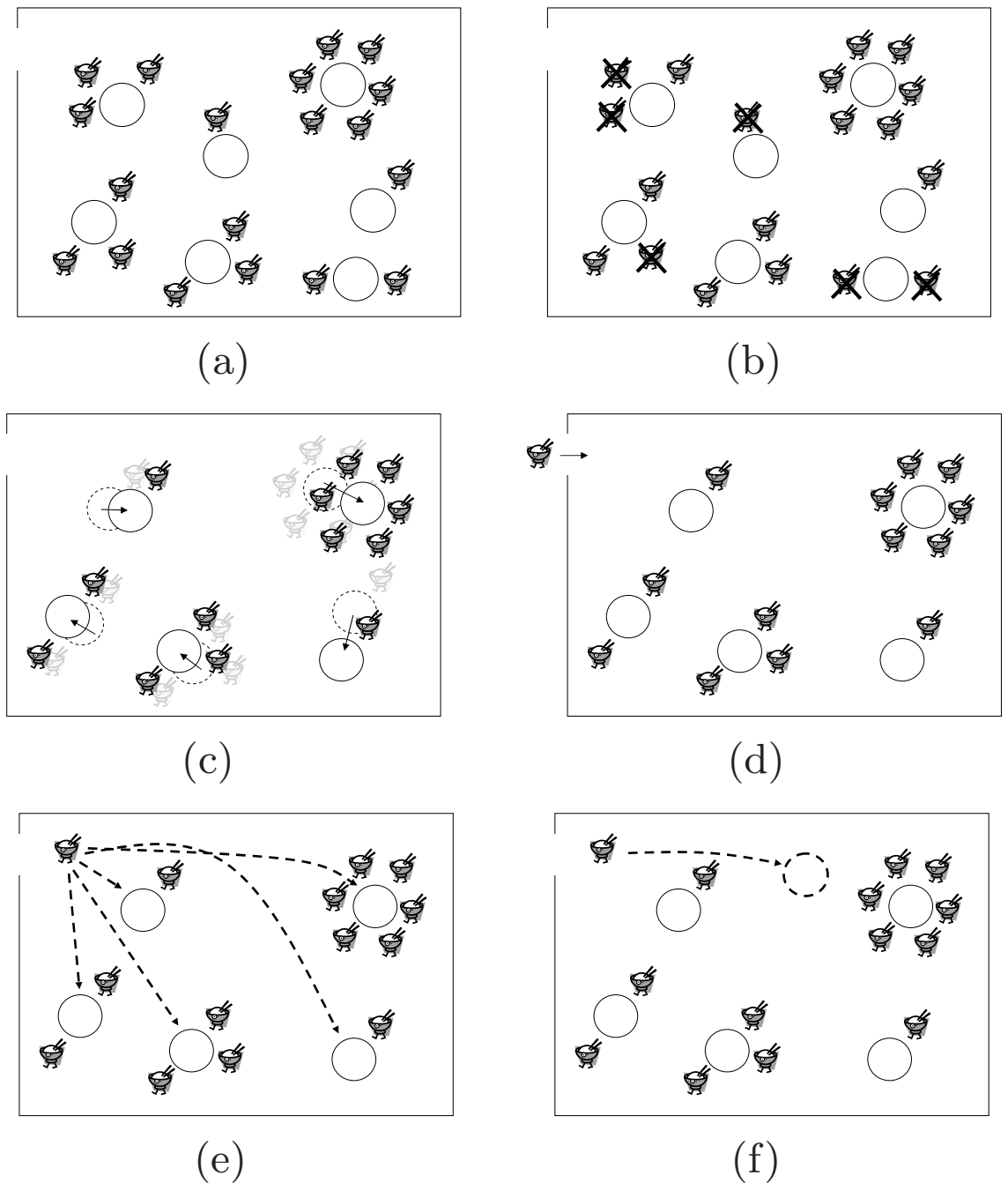

Figure 3: Illustration of the (uniform deletion) time-varying Dirichlet process. (a) At time $t$, we suppose that there are a number of customers seating at several tables. Each customer has to make a choice: either he/she remains seated at the same table (with some probability $\rho$), or he/she definitely leaves the restaurant (with probability $(1-\rho)$). (b) Once this choice has been made by each customer, it remains in the restaurant a certain number of customers. (c) Each table that is still occupied moves according to $p(U_{k,t}|U_{k,t-1})$. (d) A new customer enters the restaurant and either (e) seats at a table with a probability depending on the number of people at this table or (f) seats alone at a new table whose localization is distributed from $\mathbb{G}_0$. $n-1$ other new customers enter the restaurant, and repeat operations (d)-(f).

## 3 Bayesian Inference in Time-Varying DPM

Inference is based on the posterior distribution of the cluster assignment variables $\mathbf{c}_t$, the vectors $\mathbf{m}_{t-1}^t$ and $U_{1:K_t}$ given by $p(\mathbf{c}_{1:t}, \mathbf{m}_{1:t-1}^{1:t}, U_{1:K_t}|\mathbf{z}_{1:t})$ at time $t$. This distribution does not admit a closed-form expression and we propose to estimate them by Monte Carlo methods. We use a Sequential Monte Carlo (SMC) algorithm [6] for online inference to sample from the sequence of distributions $p(\mathbf{c}_{1:t}, \mathbf{m}_{1:t-1}^{1:t}, U_{1:t}|\mathbf{z}_{1:t})$ as $t$ increases. The algorithm relies on importance distributions denoted generically $q(\cdot)$ and is initialized with $w_0^{(i)} = N^{-1}$ and $m_0^{0\ (i)} = \varnothing$ for $i = 1, ..., N$.

---

**SMC for uniform deletion model**

---

*At time $t \geq 1$*

- For each particle $i = 1, .., N$
  - Sample $\widetilde{\mathbf{m}}_{t-1}^{t\ (i)}|\mathbf{m}_{t-1}^{t-1\ (i)} \sim \Pr(\mathbf{m}_{t-1}^t|\mathbf{m}_{t-1}^{t-1\ (i)})$
  - Sample $\widetilde{\mathbf{c}}_t^{(i)} \sim q(\mathbf{c}_t|\widetilde{\mathbf{m}}_{t-1}^{t\ (i)}, U^{(i)}_{\mathcal{I}(\widetilde{\mathbf{m}}_t^{(i)}),t-1}, z_t)$
  - For $k \in \mathcal{J}(\widetilde{\mathbf{c}}_t^{(i)}) \cap \overline{\mathcal{I}(\widetilde{\mathbf{m}}_{t-1}^{t\ (i)})}$, sample $\widetilde{U}_{k,t}^{(i)} \sim q(U_{k,t}|\{\mathbf{z}_{j,t}|\widetilde{c}_{j,t}^{(i)} = k\})$
  - For $k \in \mathcal{I}(\widetilde{\mathbf{m}}_{t-1}^{t\ (i)})$, sample $\widetilde{U}_{k,t}^{(i)} \sim \begin{cases} q(U_{k,t}|U_{k,t-1}^{(i)}, \{\mathbf{z}_{j,t}|\widetilde{c}_{j,t}^{(i)} = k\}) & \text{if } k \in \mathcal{J}(\widetilde{\mathbf{c}}_t^{(i)}) \\ p(U_{k,t}|U_{k,t-1}^{(i)}) & \text{otherwise} \end{cases}$
- For $i = 1, .., N$, update the weights

$$\begin{aligned} \widetilde{w}_t^{(i)} &\propto w_{t-1}^{(i)} \prod_{k=1}^n \frac{p(\mathbf{z}_{k,t}|\widetilde{U}^{(i)}_{\widetilde{c}_{k,t}^{(i)},t}) \Pr(\widetilde{\mathbf{c}}_t^{(i)}|\widetilde{\mathbf{m}}_{t-1}^{t\ (i)})}{q(\widetilde{\mathbf{c}}_t^{(i)}|\widetilde{\mathbf{m}}_{t-1}^{t\ (i)}, U^{(i)}_{\mathcal{I}(\widetilde{\mathbf{m}}_t^{(i)}),t-1}, \mathbf{z}_t)} \\ &\times \prod_{k\in\mathcal{I}(\widetilde{\mathbf{m}}_{t-1}^{t\ (i)})} \frac{p(\widetilde{U}_{k,t}^{(i)}|U_{k,t-1}^{(i)})}{q(\widetilde{U}_{k,t}^{(i)}|U_{k,t-1}^{(i)}, \{\mathbf{z}_{j,t}|\widetilde{c}_{j,t}^{(i)}=k\})} \\ &\times \prod_{k\in\mathcal{J}(\widetilde{\mathbf{c}}_t^{(i)})\cap\overline{\mathcal{I}(\widetilde{\mathbf{m}}_{t-1}^{t\ (i)})}} \frac{\mathbb{G}_0(\widetilde{U}_{k,t}^{(i)})}{q(\widetilde{U}_{k,t}^{(i)}|\{\mathbf{z}_{j,t}|\widetilde{c}_{j,t}^{(i)}=k\})} \end{aligned} \quad (9)$$

with $\sum_{i=1}^N \widetilde{w}_t^{(i)} = 1$.

- Resampling. Compute $N_{\text{eff}} = \left[\sum \left(\widetilde{w}_t^{(i)}\right)^2\right]^{-1}$. If $N_{\text{eff}} \leq N/2$, duplicate the particles with large weights and remove the particles with small weights, resulting in a new set of particles denoted $\cdot_t^{(i)}$ with weights $w_t^{(i)} = 1/N$. Otherwise, rename the particles and weights by removing the $\widetilde{\cdot}$.

---

To perform batch inference using MCMC, we use a different parametrization which allows us to design more efficient moves. The vectors $\mathbf{m}_{t-1}^t$ are replaced by the death times of allocation variables $c_{k,t}$, denoted $d_{k,t}$, $k = 1, \ldots, n$, $t = 1, \ldots, T$ where $d_{k,t} \geq t$. From $\{c_t\}_{t=1,\ldots,t'}$ and $\{d_t\}_{t=1,\ldots,t'}$ one can reconstruct the vector $\mathbf{m}_{t-1}^t$. At each time step $t$, we sample successively $\mathbf{c}_t$, $\mathbf{d}_t$ and $U_t$. The usual MCMC techniques for DPMs [15] can be adapted to our model. For each iteration $i = 1, \ldots, N$, the algorithm is the following.

---

**MCMC for uniform deletion model**

---

- For $t = 1, \ldots, T$
- For $k = 1, \ldots, n$, sample $c_{k,t} \sim \Pr(c_{k,t}|\mathbf{c}_{-k,t}, \mathbf{c}_{-t}, \mathbf{d}_{1:T}, U_t, \mathbf{z}_{1:T})$.
  - If $c_{k,t}$ takes a new value, update $\mathbf{m}_{t+1:d(k,t)}^{t+1:d(k,t)}$ and $\mathbf{m}_{t:d(k,t)}^{t+1:d(k,t)}$ and sample $\mathbf{U}_{1:d_{k,t}} \sim p(\mathbf{U}_{c_{k,t},1:d_{k,t}}|\mathbf{z}_{k,t}, c_{k,t})$. Let $c_{k,t}^{old}$ be the old value.

For $u = t+1,\ldots,d_{k,t}$, if $m^u_{c^{old}_{k,t},u-1} = 0$ and $m^u_{c^{old}_{k,t},u} > 0$, then create a new cluster and relabel $c^{old}_{k,t}$ for $u' = u,\ldots,min\{u'' > u | m^{u''}_{c^{old}_{k,t},u''-1} = 0$ and $m^{u''}_{c^{old}_{k,t},u''} > 0\} - 1$.

- For $k = 1,\ldots,n$, sample $d_{k,t} \sim \Pr(d_{k,t}|\mathbf{d}_{-k,t},\mathbf{d}_{-t},\mathbf{c}_{1:T})$. Let $d^{old}_{k,t}$ be the old value. For $u = \min(d^{old}_{k,t},d_{k,t})+1,\ldots,\max(d^{old}_{k,t},d_{k,t})$, update $m^u_{c_{k,t},u-1}$ and $m^u_{c_{k,t},u}$ by increasing or decreasing its value by 1.

- If $d_{k,t} > d^m = \max(\{d_{a,u}|c_{a,u} = c_{k,t}\})$, sample $\mathbf{U}_{d^m+1:d_{k,t}} \sim p(\mathbf{U}_{c_{k,t},d^m+1:d_{k,t}}|U_{c_{k,t},d^m})$.

- If $d_{k,t} < d^{old}_{k,t}$, for $u = d_{k,t}+1,\ldots,d^{old}_{k,t}$, if $\mathbf{m}^u_{c_{k,t},u-1} = 0$ and $\mathbf{m}^u_{c_{k,t},u} > 0$, then create a new cluster and relabel $c_{k,t}$ for $u' = u,\ldots,min\{u'' > u|\mathbf{m}^{u''}_{c_{k,t},u''-1} = 0$ and $\mathbf{m}^{u''}_{c_{k,t},u''} > 0\} - 1$.

- For $j \in \mathcal{I}(\mathbf{m}^t_t)$, sample $U_{j,t} \sim p(U_{j,t}|U_{j,t-1},U_{j,t+1},\mathbf{c}_t)$.

# 4 Applications

## 4.1 Sequential time-varying density estimation

We consider the synthetic problem of estimating sequentially a sequence of time-varying densities $F_t$ on the real line using the observations $\mathbf{z}_t$. We assume the sequence of observations $\mathbf{z}_t$ (where $n = 1$) follows the TVDP defined in this article with both uniform and size-biased deletion, a Gaussian mixed density and Normal-inverse Gamma base distribution. To keep the presentation simple, we assume here that the hyperparameters of the base distribution are assumed fixed and known $\mu_0 = 0$, $\kappa_0 = 0.1$, $\nu_0 = 2$ and $\Lambda_0 = 1$. The DP scale parameter is $\theta = 3$. Instead of fixing $\rho$, we assume it is time-varying with $p(\rho_t|\rho_{t-1}) = \mathcal{B}(a_\rho, a_\rho\frac{1-\rho_{t-1}}{\rho_{t-1}})$ where $a_\rho = 1000$, such that $\mathbb{E}[\rho_t|\rho_{t-1}] = \rho_{t-1}$ and $var(\rho_t|\rho_{t-1}) = \frac{\rho^2_{t-1}(1-\rho_{t-1})}{a_\rho+\rho_{t-1}}$. Note that the resulting model is still first order stationary. We select a mixture of uniform and size-biased deletions with $\alpha = 0.98$. The observations $\mathbf{z}_t$ (one at each time step) are generated for $t = 1,\ldots,1000$ from a sequence of mixtures of normal distributions, see Fig. 4. Abrupt changes occur at times $t = 301$ and $t = 601$ where modes of the true density appear/disappear whereas the mode moves smoothly from 0 to $-1.5$ between $t = 701$ and $t = 850$. For illustration purposes, we compute the average number of alive allocation variables $N_{t|t}$ as follows

$$N_{t|t} = \mathbb{E}\left[\sum_{k=1}^{K_t} m^t_{k,t} \middle| \mathbf{z}_{1:t}\right] \tag{10}$$

An SMC algorithm is implemented with 1000 particles. In Fig. 4, we display the filtered density estimate $F_{t|t} = \mathbb{E}[F_t|\mathbf{z}_{1:t}]$ which manages to track the slow and abrupt changes of the true density. The evolution of the SMC estimates of $N_{t|t}$ and $\rho_{t|t} = \mathbb{E}[\rho_t|\mathbf{z}_{1:t}]$ are given in Fig. 5. We see that the model adapts to $F_t$ quickly by also estimating $\rho_t$: $\rho_t$ suddenly decreases at times $t = 300$ where the modes of the density suddenly change. $N_{t|t}$ follows a similar evolution. Whenever $F_t$ does not evolve, the algorithm uses as many previously collected observations as possible to estimate the density by letting $N_{t|t}$ increase. When $F_t$ changes abruptly, $N_{t|t}$ decreases abruptly too and the model quickly gets rid of the old clusters. Moreover, by using a size-biased deletion procedure, we allow the algorithm to delete only some modes of the density while keeping the remaining allocation variables alive. This is illustrated at $t = 600$ where the two minor modes disappear while the main one is preserved.

## 4.2 Dynamic topic model clustering

Topic models are a subject of great interest in machine learning due to the increasing number of digital documents and the need for an automated organization of information. We represent text corpora as bag-of-words and represent documents as vectors containing word frequencies, disregarding their order [4]. We are here interested in modeling time-varying topics; see [3], [19] for recent work on the subject.

We consider that, at each time $t$, we have an infinite number of topics. Each topic is represented by a vector of size $K$ ($K$ being the size of the vocabulary) representing the relative frequencies of each vocabulary word for that topic. Let $\mathbb{G}_t \sim DP(\theta,\mathbb{G}_0)$ be the prior distribution of the topics at time $t$, where $\mathbb{G}_0 = \mathcal{D}(\frac{\theta_V}{K},\ldots,\frac{\theta_V}{K})$ is a Dirichlet distribution and $\theta_V$ a fixed hyperparameter, $\theta_V = 0.5$.

We suppose that we associate a topic $\mathbf{y}_{k,t}$ to each word $w_{k,t}$ such that

$$\mathbf{y}_{k,t} \sim \mathbb{G}_t,\ w_{k,t} \sim \mathcal{M}(\mathbf{y}_{k,t})$$

where $\mathcal{M}$ is the multinomial distribution.

We analyze here the titles of the Proceedings of the National Academy of Sciences (PNAS) from 1915 to 2000[1]. This corpus has been filtered by removing stop words and words occurring less than 70 times in the whole corpus. After this preprocessing step, 283425 words and a vocabulary size $K = 1021$ remain. The time index corresponds to the year of publication, and the number of words for each year is varying. We assume that the words $w_{k,t}$ are distributed according to a time-varying DP based on an uniform deletion procedure with $\rho = 0.4$. The topics are assumed fixed and we integrate them out in the MCMC algorithm which is run for 2000 iterations.

[1] Downloadable at http://www.cs.toronto.edu/~roweis/data.html

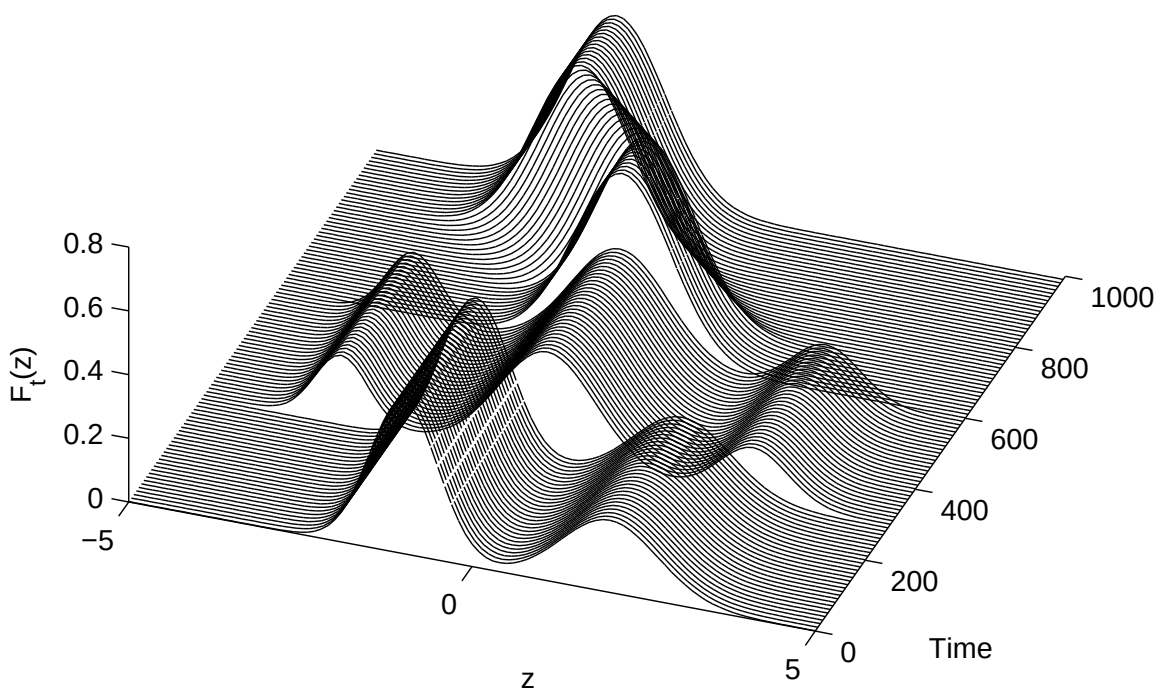

(a) True density

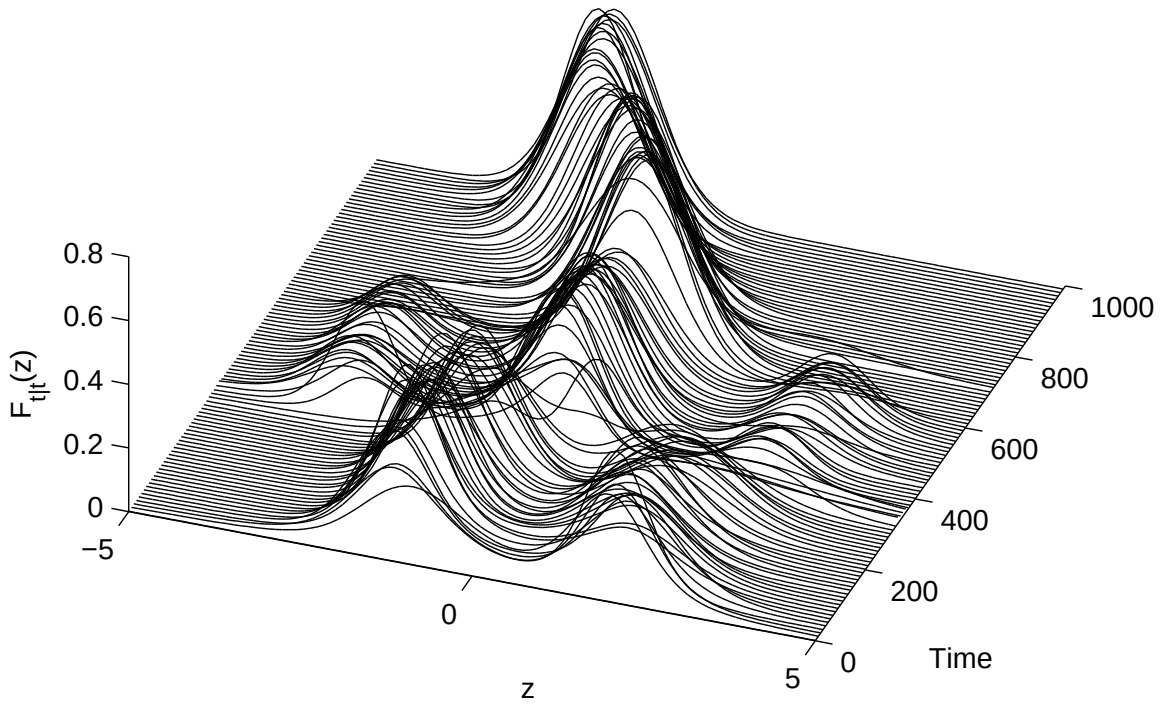

(b) Estimated density

Figure 4: (a) True density and (b) estimated density. Abrupt changes occur at times $t = 301$ and $t = 601$. The mode of the density evolves smoothly between times $t = 700$ and $t = 850$.

The resulting posterior probability for some words and some topics are shown in Fig. 6. The model is able to capture the evolution of topics over time and allows for the appearance/disappearance of topics, e.g. topics 3 and 4.

### Discussion

There are several potential methodological extensions to this work. First, although the details are omitted here, it is actually possible to extend this work of the class of Pitman-Yor processes. Second, it would be interesting to develop models allowing clusters to merge or split over time.

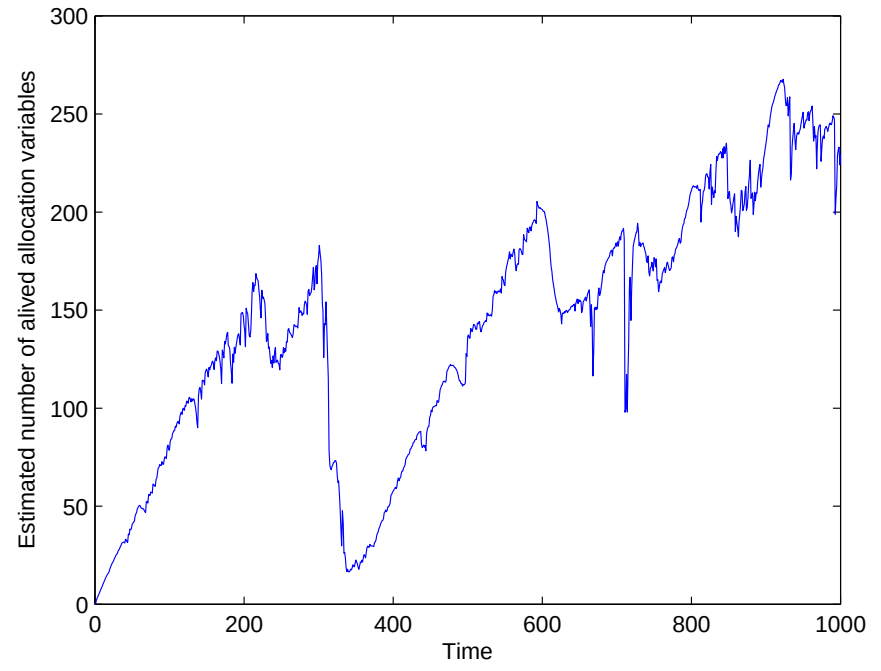

(a)

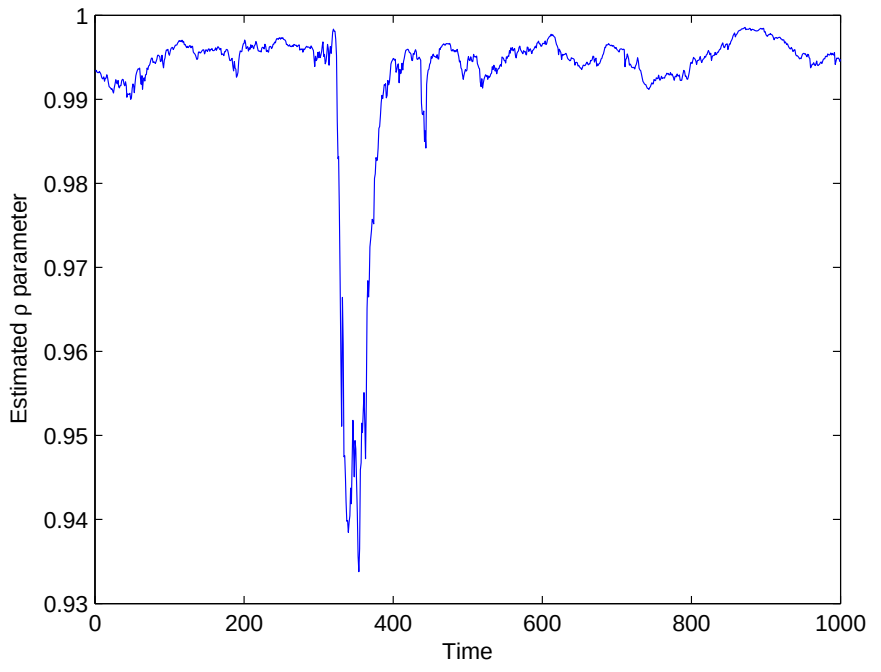

(b)

Figure 5: (a) Time evolution of the average number of alive clusters. (b) Time evolution of $\rho_{t|t}$.

### Acknowledgements

The first author is grateful to DGA for its financial support.

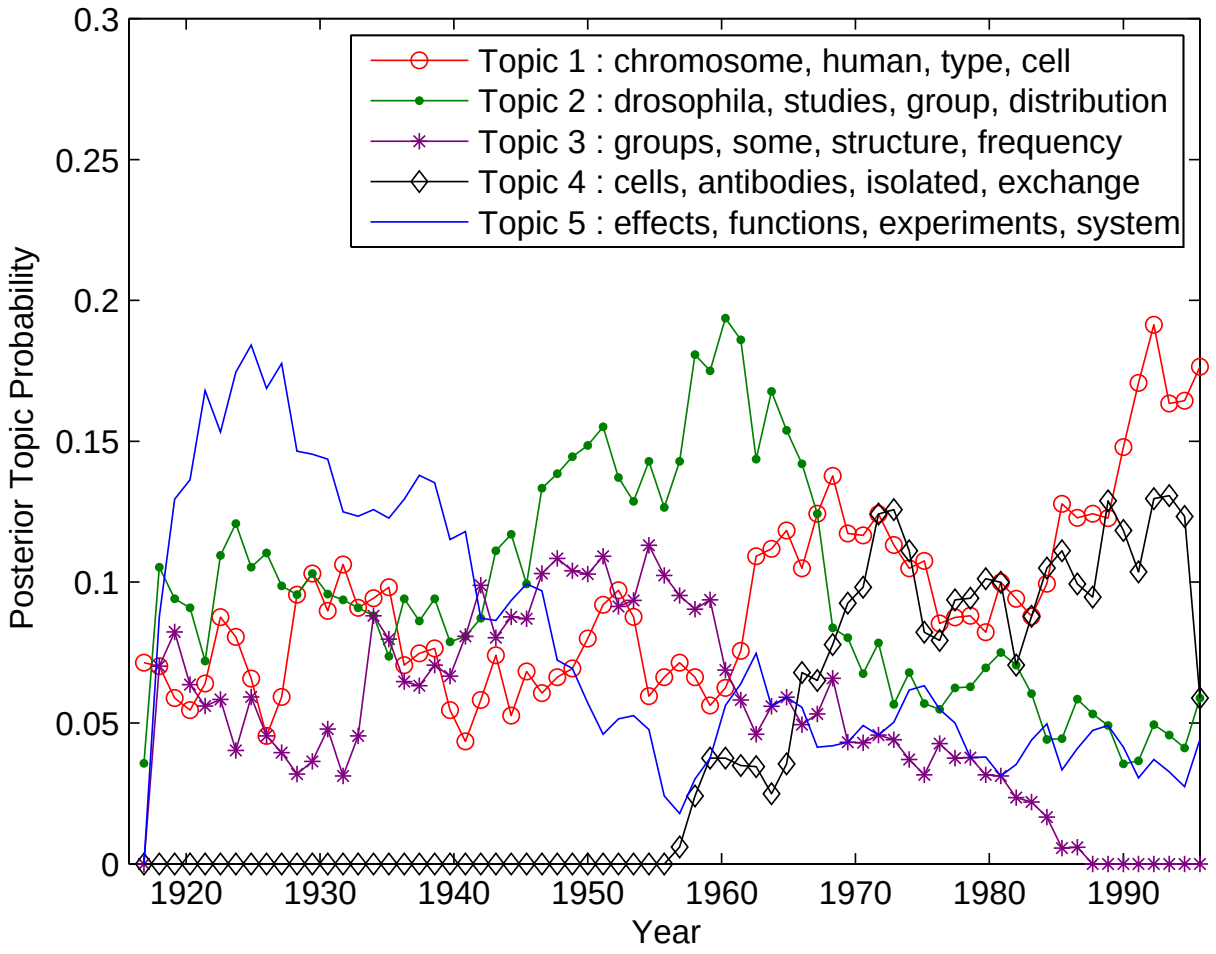

(a)

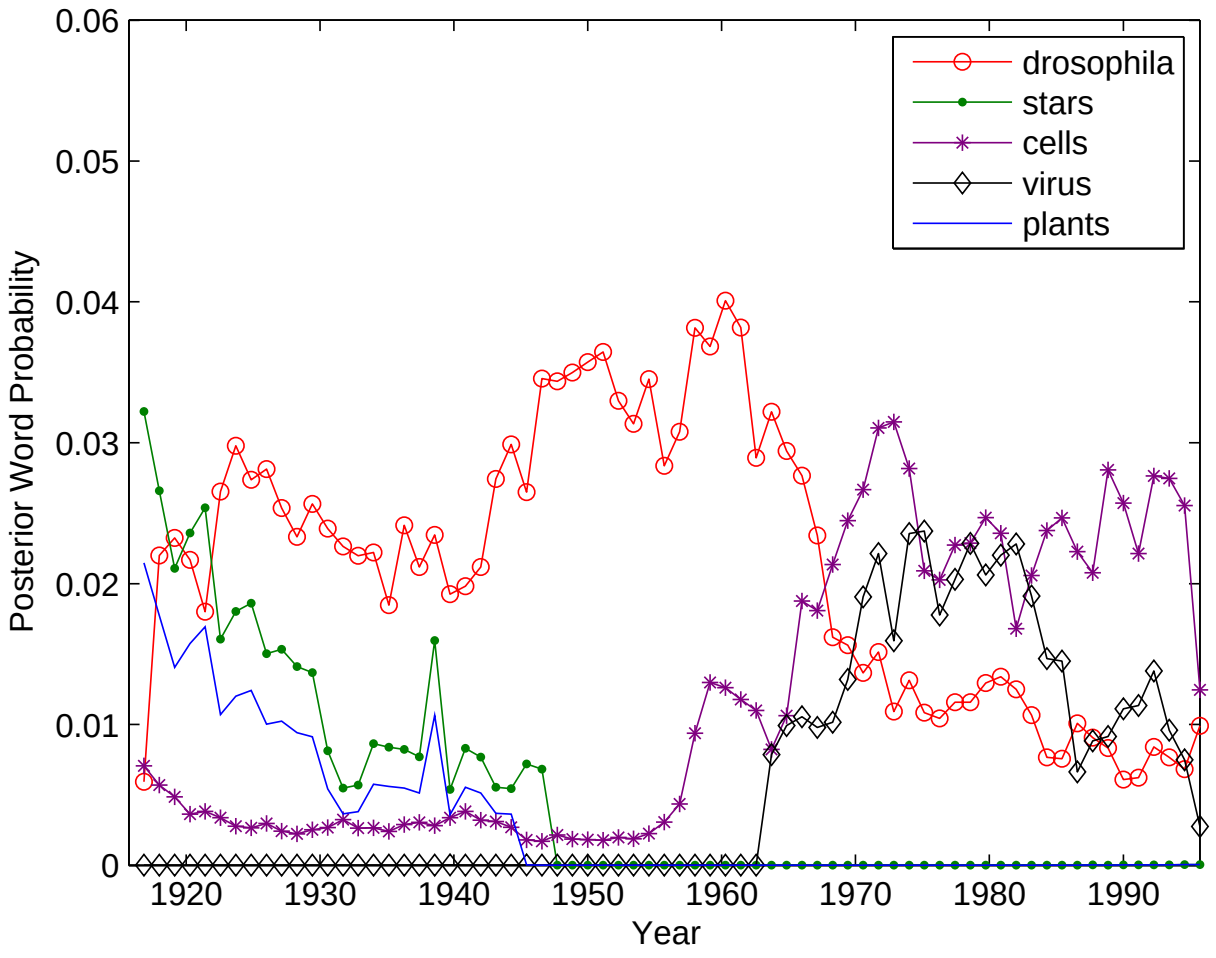

(b)

Figure 6: (a) Posterior probability for some topics in function of time. (b) Posterior probability for some words in function of time.